# Title:
# Understanding inhomogeneous crystallization dynamics of phase-change materials in the vicinity of metallic nanoantennas

**Author(s), and Corresponding Author(s)*:**

Luis Schüler*, Lukas Conrads, Yingfan Chen, Lina Jäckering, Sebastian Meyer, Matthias Wuttig, Thomas Taubner, Dmitry N. Chigrin*

*Luis Schüler and Lukas Conrads contributed equally.*

* Email: schueler@physik.rwth-aachen.de
* Email: chigrin@physik.rwth-aachen.de

**Affiliations:**

AMO GmbH, Otto-Blumenthal-Straße 25, 52074 Aachen, Germany
  L. Schüler, D. N. Chigrin

I. Institute of Physics (IA), RWTH Aachen University, Sommerfeldstraße 14, 52074 Aachen, Germany
  L. Schüler, L. Conrads, Y. Chen, L. Jäckering, S. Meyer, M. Wuttig, T. Taubner, D. N. Chigrin

**Abstract:**

Optical metasurfaces composed of metallic or dielectric scatterers (meta-atoms) promise a powerful way of tailoring light-matter interactions. Phase-change materials (PCMs) are prime candidates for non-volatile resonance tuning of metasurfaces based on a refractive index change. Precise resonance control can be achieved by locally applying laser pulses to crystallize a PCM, modifying the dielectric surrounding of meta-atoms. However, the complex crystallization kinetics of PCMs in the vicinity of metallic meta-atoms have not been studied yet. Here, we experimentally investigate metallic dimer antennas on top of the PCM $Ge_3Sb_2Te_6$ and address these nanoantennas with laser pulses. Our study reveals inhomogeneous crystallization caused by the absorption and heat conduction of the metallic nanoantennas. A self-consistent multiphysics model, including electromagnetic, thermal, and phase-transition processes, is employed to simulate the crystallization and understand the resulting resonance shift of the antennas. This model enables the optimization of the laser parameters and the geometry of the meta-atoms to achieve an optimal crystallization pattern and resonance shift. Our work paves the way towards complex antenna geometries optimized for local addressing of PCMs to achieve sophisticated crystallization patterns, enabling on-demand programming of individual nanoantennas within metasurfaces.

**Introduction:**

Metasurfaces composed of subwavelength metallic or dielectric scatterers have been widely investigated in the past years due to their ability to tailor light-matter interaction at will by abrupt manipulation of phase, amplitude and polarization of the scattered light.[1]–[3] The realized applications range from beam steering[4] and lensing[5] to more complex phenomena such as holography[6].

In general, the functionality and operating wavelength of the designed metasurfaces are fixed after fabrication and cannot easily be altered. Still, several approaches to enable dynamic modulation of the metasurfaces arose[7], such as using stretchable substrates[8], injecting charge carriers[9],[10], or exploiting the insulator-to-metal transition of $VO_2$.[11] However, all these approaches are non-reversible or volatile since the effect only persists as long as the external stimulus is applied.

For non-volatile resonance tuning of metallic nanoantennas, chalcogenide-based phase-change materials (PCMs) have evolved as a promising platform.[12],[13] They generally feature two (meta-)stable phases, the amorphous and the crystalline one, which differ significantly in their optical and electrical properties.[14] This difference is attributed to a change in bonding upon phase-change. While the atoms in the amorphous phase are covalently bonded, the atoms are metavalently bonded in the crystalline phase for most PCMs.[15]–[19] Reversible switching between both phases is achieved by heating the PCM either locally with a laser or by applying electrical pulses, or globally on a heating plate. The fast non-volatile switching combined with the strong contrast in refractive index between both phases has been exploited for active nanophotonic components such as switchable lenses and beam steerers[20]–[22], tunable absorbers[23], and holography,[24]–[26] by switching the entire metasurface at once.

In contrast, local addressing of individual meta-atoms for fully programmable infrared metasurfaces has been demonstrated by precisely crystallizing and reamorphizing the PCM around single metallic nanoantennas with visible laser pulses.[27] The same approach has been transferred to more complex antenna shapes such as split-ring resonators, independently modifying their magnetic and electric dipole resonances by addressing the resonators at different positions.[27],[28] It was found that the metallic nanoantennas in vicinity of the PCMs lead to inhomogeneous and time-dependent light absorption and heat conduction within the layer stack. These effects play a pivotal role for the observed crystallization patterns and lead to non-uniform spatial phase transitions of the PCM. Hence, a self-consistent multiphysics description of the light absorption, heat conduction and temperature evolution as well as phase-transition kinetics is needed for accurate theoretical description of the experimental setup and to understand the underlying physical phenomena occurring during the phase-change.[29]

Different models of the crystallization of PCMs have been applied in literature: Ab initio molecular dynamics simulations offer unrivaled atomistic insights into the phase transition processes but are computationally limited to small time and length scales.[30]–[32] In contrast, effective medium methods using the Johnson-Mehl-Avrami-Kolmogorov (JMAK)[33]–[37] formalism are ideal for fast macroscopic simulations but lack spatial resolution or time-dependent effects. An improved model based on rate equations for nucleation and growth can incorporate time-dependent and transient effects at an increased computational cost.[38] However, these approaches are not suited for multiphysics models aiming at a comprehensive and meta-atom-scale simulation of phase change processes including time- and location-dependent optical or electrical effects.

Instead, for example, models simulating the crystallization using classical nucleation theory[39]–[41] or Gillespie-type cellular automata approaches[42],[43] have been successfully employed. A popular alternative are phase-field models[29],[44]–[48], as they feature a reasonable trade-off between computation times and the resolution of physical details. Using continuous fields, they can also capture complex phenomena such as anisotropy. However, a detailed study about the crystallization behavior of PCMs in vicinity of metallic nanoantennas upon laser irradiation with focus on the different underlying mechanisms has not been shown yet.

Here, we experimentally investigate the laser-induced crystallization behavior of the PCM $Ge_3Sb_2Te_6$ (GST) in the vicinity of metallic nanoantennas and leverage multiphysics simulations for the underlying physical processes. In particular, aluminum dimer antennas are placed on top of the PCM GST. Subsequent local crystallization of the PCM at the center and the edge of the dimer antennas reveals strongly inhomogeneous crystallization that opposes the naively expected uniform crystallization according to the elliptical intensity profile of the incident laser. The measured infrared reflectance spectra are compared to simulated spectra of a simple cylindrical crystallization volume and simulated spectra accounting for the inhomogeneous crystallization. We apply three-dimensional multiphysics simulations to reveal the power losses within the materials upon laser irradiation, translate the simulated power losses to a temperature distribution within the PCM and finally apply a crystallization kinetics model to obtain the crystallization pattern of the PCM around the nanoantennas. The results underline the need for accurate description of the modelled structure to obtain sufficient agreement between simulation and experiment. Furthermore, our results demonstrate how the interaction with metallic nanostructures and parameters like the polarization direction of the laser can significantly influence the crystallization behavior of PCMs, resulting in resonance shifts that are difficult to predict.

**Main Text:**
Optical crystallization of PCM thin films throughout the whole film has been extensively studied in literature.[27],[49],[50] Naively, a cylindrical crystallized volume, laterally resembling the intensity distribution of the switching laser employed, would be expected as schematically displayed in **Figure 1a**. However, this simplified picture does not represent reality. In fact, the thermal properties of the adjacent layers, i.e., the heat confinement at the top of the structure due to the small heat conductivity of air, lead to a finite crystallization depth as well as slight deviations from the elliptical spot due to the nucleation probability of the PCMs.[27],[50] Taking now also metallic nanoantennas, such as split-ring resonators, in the vicinity of the PCM into account, the crystallization behavior changes drastically due to interaction between the laser and the metallic structures, and the large thermal conductivity of the metal.[28] In the following, we will study the influence of plasmonic nanoantennas on the crystallization behavior of PCM layers.

The aluminum antennas investigated are fabricated on a silicon substrate covered with a 50 nm thin layer of amorphous $Ge_3Sb_2Te_6$ (GST), followed by a 70 nm thin capping layer of ZnS:SiO$_2$ (see **Methods** for more details). As sketched in **Figure 1b**, the dimer antennas are composed of two rod antennas separated by a 200 nm gap, each with a length of 1000 nm, a width of 300 nm, and a height of 42 nm. The crystallization of the PCM around the antennas is induced by laser irradiation with a wavelength of 660 nm. The laser can be characterized as an elliptical Gaussian beam that is horizontally polarized, along the major axis of the ellipse. Upon crystallization, the real part of the permittivity of GST changes from 12 to 36 in the infrared spectral range, c.f. **Figure 1c**.

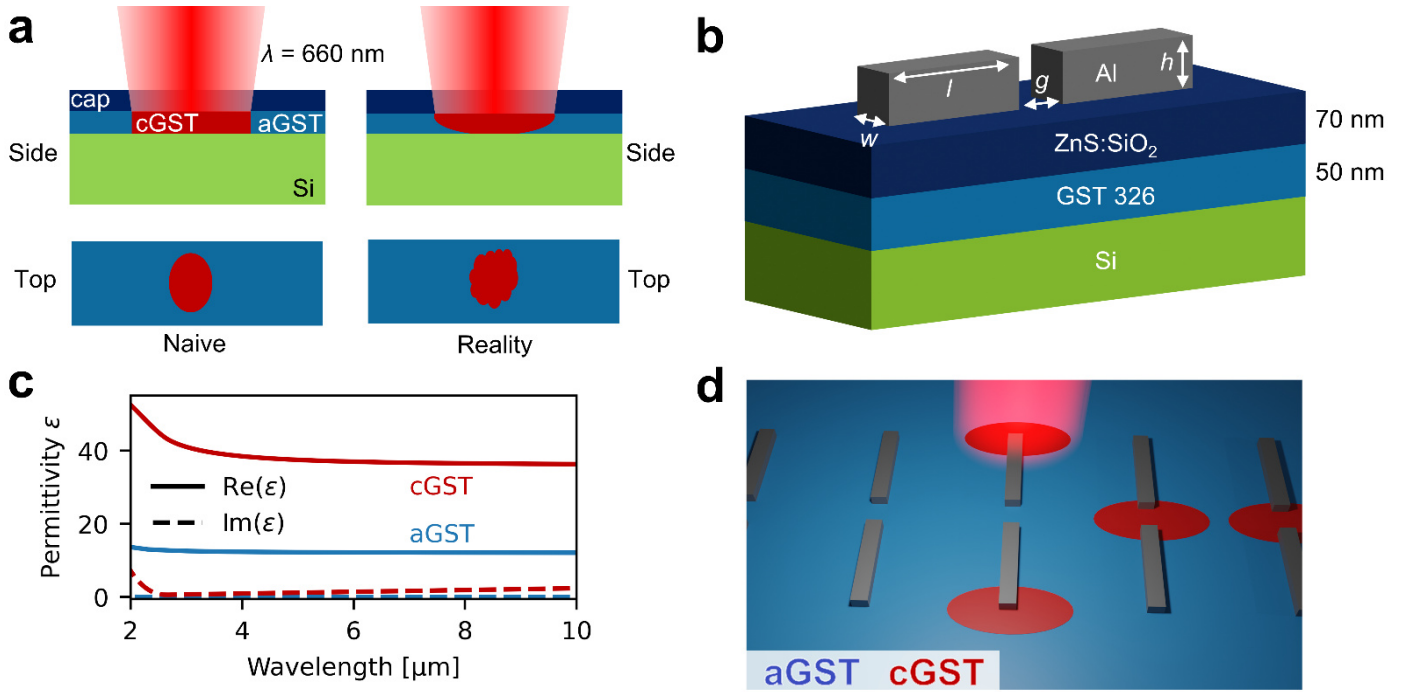

**Figure 1: Crystallization at a metallic structure - concept. a)** Schematic sketch of naive crystallization of GST (left) where the crystallized volume is given by an elliptical cylinder (due to an elliptical intensity distribution) in contrast to the reality (right) where a finite crystallization depth in addition to slight irregularities can be observed. **b)** Aluminum dimer antennas with geometric parameters of $w$ = 300 nm, $l$ = 1000 nm, $g$ = 200 nm, and $h$ = 42 nm positioned on 50 nm thin amorphous GST and 70 nm thin capping. **c)** Dielectric permittivity of amorphous (blue) and crystalline (red) GST. **d)** Sketch of laser-induced crystallization of GST with dimer antennas on top, either at the ends of the dimer antennas or at the center to study the influence of the laser spot position for the observable crystallization.

Here, we distinguish two different positions to locally address the GST below the nanoantennas, see **Figure 1d:** The dimer antennas can be locally addressed either at the antenna edges, or at the center of the dimer antennas. The position of the crystallized region plays a crucial role for antenna resonance tuning. In particular, the refractive index at the hotspots of the antenna, referring to enhanced electric fields upon resonantly exciting the antenna, has the most significant effect on the resonance position.[28] The metallic nanoantennas strongly alter the observed crystallization behavior. In the following, we will show how exactly the crystallization is affected by the nanoantennas and how the complex crystallization affects e.g. reflectance spectra of these antennas. Then, by disentangling the different processes involved and with the use of multiphysics simulations, we explain the causes of the complex crystallization.

Light microscope images of the locally crystallized GST for center addressing and edge addressing are shown in **Figure 2a**. Exemplarily, a laser power of 15.6 mW with 500 ns pulse duration was chosen. Despite the well-defined elliptical intensity distribution of the switching laser, the crystallization pattern observed displays a clear sub-structure which is attributed to the influence of the antenna structures. Moreover, the crystallization pattern for the edge-addressed antennas shows a mushroom-like structure. Since optical light microscopy images are diffraction limited and thus cannot resolve fine structures in the crystallization pattern, we employ scattering-type scanning near-field optical microscopy (s-SNOM) as a tool for imaging the crystallization pattern with a resolution down to several nanometers. Sub-wavelength resolution is achieved by probing the sample with highly localized near fields of a sharp tip, which is illuminated with infrared laser light with wavelengths far off from the resonance of the antennas (see **Methods** for more details). The corresponding recorded s-SNOM amplitude image in the third demodulation order (see **Figure 2b**) reveals sharp bright regions corresponding to cGST around the dimer antennas. Multiphysics simulations with a laser beam matching the experimental parameters are performed by taking the optical and thermal properties of the materials employed into account and self-consistently calculating the crystallization behavior of GST. The simulation time was set to 500 ns, matching the experimental pulse length of the laser, with 50 steps of 10 ns duration each. For each time step of the simulation, the power loss within the layers is first simulated with an electromagnetic solver, subsequently transformed into the time-dependent temperature distribution using a thermal solver, and finally used as an input for a crystallization kinetics model to retrieve the crystallization pattern (see Methods for more details). As shown in **Figure 2c**, the calculated crystallized volume in top view features strongly inhomogeneous patterns that barely resemble the original shape of the incident laser in the focal plane and depend on the position of the incident laser. The inhomogeneous crystallization patterns from the simulations, i.e., butterfly-pattern for center-addressed and mushroom-pattern for edge-addressed dimers, match the ones observed in the light microscope and s-SNOM images.

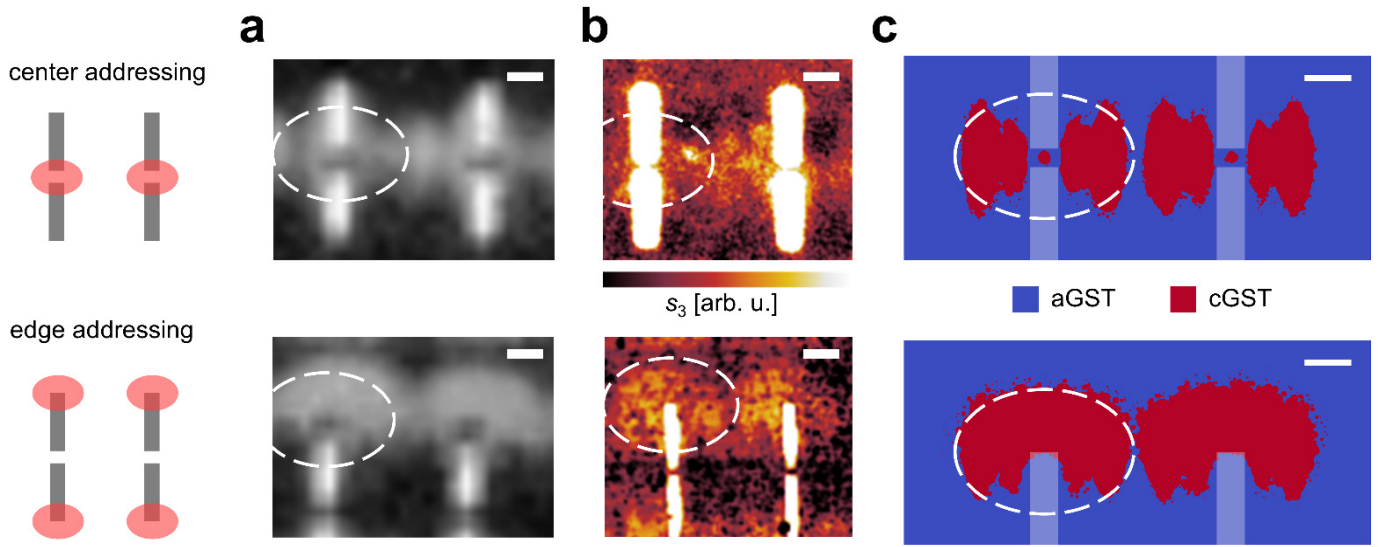

**Figure 2: Laser-induced crystallization of GST at the dimer antennas.** The laser pulses (red ellipses) are applied in the center or at the edges of the dimer antennas. The scale bars are 0.5 μm. The white-dashed ellipses indicate the approximate size of a crystallized spot without antennas. **a), b)** Light microscopy images (**a**) and the third-order demodulated amplitude from scattering-type scanning near-field optical microscopy (**b**) of dimer antennas after laser crystallization display spatially inhomogeneous crystallization. **c)** Simulated crystallization of the GST layer, seen from above. The location of the antennas on top of the GST is indicated by semi-transparent rectangles. For center-addressed antennas, a butterfly pattern occurs, while for edge-addressed antennas, a mushroom pattern is visible.

To highlight the effect of the inhomogeneous crystallization, we acquired infrared reflectance spectra (see Methods) for antennas on amorphous GST and for antennas where the GST was crystallized by locally addressing it with the laser, see **Figure 3a**. The antennas were fabricated in a 20 x 20 µm² array with periods of 1.5 µm and 3 µm in *x*- and *y*-direction, respectively. The reflectance curves display a pronounced peak around 5.5 µm, corresponding to the electric dipole resonance of the dimer antennas (see field simulations in **Supplementary Note 1**). If GST is in the amorphous phase (blue curve), the electric dipole resonance is located at about 5.5 µm. Upon laser crystallization of the dimer antennas at the center, the electric dipole resonance wavelength surprisingly does *not* shift in the experiment. However, this laser power results in a crystallized elliptical spot of about 2.0 μm (diameter along major axis) on amorphous GST without antennas (c.f. **Supplementary Note 2**). If the experimentally measured size of the crystallized spot is included in numerical simulations (c.f. **Figure 3b**) and the crystallization is simulated as an elliptically shaped cylinder (see inset for top view), the electric dipole resonance shifts to 6 μm (red dotted curve). The sharp features, mainly between 6 μm and 7 μm, correspond to grating resonances caused by the antennas being placed periodically in an array. They are strongly suppressed in the experiment due to several factors: first, the use of a Schwarzschild objective featuring an angular range, providing an averaging effect; second, the crystallization being different at each antenna due to its stochastic nature; and third, the antennas possibly deviating slightly in their geometry. To resolve this discrepancy, we perform multiphysics simulations to retrieve the crystallization volume (see inset for the top view of the crystallization volume) and utilize this crystallization volume around the nanoantennas for simulated reflectance spectra (red dashed curve). Now, the electric dipole resonance wavelength is well reproduced within the simulations. We attribute the negligible resonance shift to the inhomogeneous crystallization pattern, specifically the GST not being crystallized directly at or below the antennas. Especially crystallization directly at the antenna ends would lead to a resonance shift, as the electric field of the resonantly excited antenna is highest there.

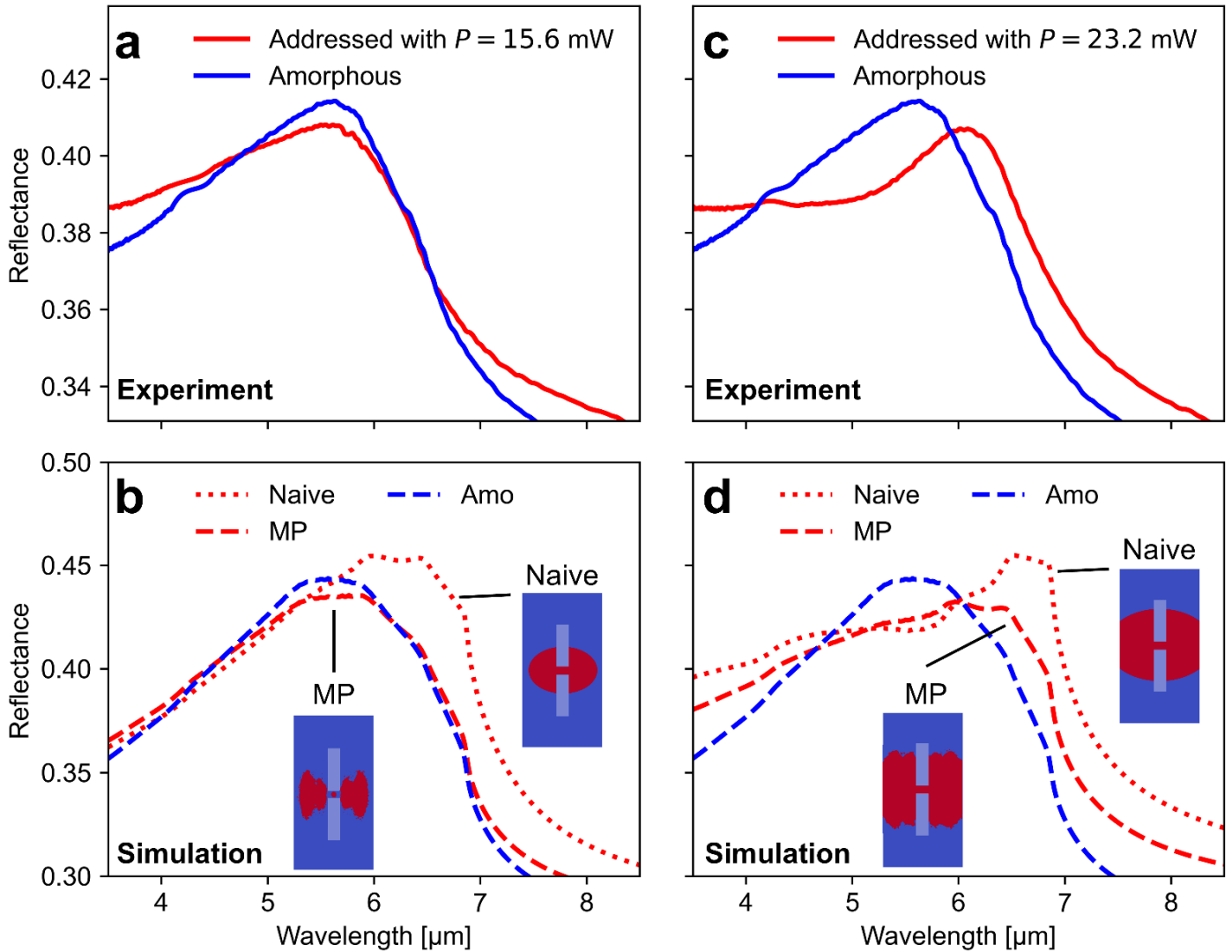

**Figure 3: Far-field reflectance spectra of locally addressed dimer antennas compared to multiphysics simulations. a)** Measured reflectance (solid lines) of the dimer antennas above amorphous GST (blue) and GST locally crystallized with 15.6 mW (red), surprisingly featuring the same resonance wavelength. **b)** If a simple elliptical cylinder is assumed as the crystallized volume (naive simulation), the resonance redshifts. In contrast, a multiphysics simulation reveals strongly inhomogeneous crystallization, resulting in no shift of the resonance. **c)** If the GST below the dimer antennas is locally addressed with a higher laser power of 23.2 mW, the resonance redshifts. **d)** The crystallization becomes more homogeneous and the difference between a simple naive and multiphysics simulation decreases, leading to a similar shift in the antenna resonance wavelength for both simulation procedures.

Similarly, we perform the comparison with locally crystallized spots using larger laser powers of 23.2 mW as shown in **Figure 3c**. If the applied switching power is increased, the expected crystallized GST spot length is around 2.8 μm. Now, the electric dipole resonance (peak in the reflectance) in the experiment shifts to 6 μm (c.f. red solid curve). The simulated reflectance spectra for the naive and multiphysics simulation are displayed in **Figure 3d**. While the naive simulation for an elliptically shaped crystalline GST cylinder (red dotted curve) is very similar to the reflectance curve for smaller laser powers, the reflectance curve based on multiphysics simulations (red dashed curve) is shifted to a higher wavelength. Since the GST is now crystallized in the center of the dimer antenna, specifically at the ends of the individual antennas, for both the naive approach and the multiphysics simulations, i.e., there exist regions where crystallization is present below the antennas when viewed from above, the simulated electric dipole resonance wavelength for both approaches matches with the experimental resonance. Although the crystallization obtained from multiphysics simulations now closely resembles the naive elliptical crystallization, the naive approach still predicts a larger resonance wavelength. The discrepancy can be explained by a finite crystallization depth present in the multiphysics simulations, while the naive approach assumes the PCM layer to be fully crystallized. A similar depth-dependence of the resonance wavelength has been observed before by Michel et al.[27], where a continuous dependence was found between the resonance shift of metallic nanoantennas and the crystallization depth of a GST layer placed above.

Our results show how the antenna resonance is mainly influenced by crystallization in a small region at the antenna ends, where the resonant dimer antenna features the highest electric fields (c.f. Supplementary Note 1). Crystallization next to the antennas affects the spectra only minimally and the resonance position does not change. Consequently, self-consistent multiphysics simulations with high spatial resolution are required to accurately reproduce experimental spectra of locally addressed antennas, especially if a small laser power is applied. In the following, we investigate the origin of the inhomogeneous crystallization step by step by considering the different electromagnetic and thermodynamic quantities, as well the polarization of the switching laser.

To find the physical origins of the inhomogeneous crystallization behavior observed, we first investigate electromagnetic fields within the structure, i.e., the electric field distribution and the power loss density. The electric field after laser irradiation is extracted 5 nm above the capping layer. If the dimer antennas are addressed by the horizontally polarized laser at the center, the real part of the *z*-component (out-of-plane with respect to the viewing plane) of the electric field $E_z$ (see **Supplementary Note 3** for details on the locations of the cross sections) shown in **Figure 4a** features alternating positive and negative values in *x*-direction. The surface charge density (see **Figure 4b**) alternates in *x*-direction across the width of the antenna and is constant in *y*-direction except for a slight decrease in absolute value. We attribute this behavior to the oscillating charge carriers within the antenna.

The resulting simulated power loss density at the top surface of the GST as displayed in **Figure 4c**. Several pronounced regions of enhanced power loss are visible: First, the power loss is enhanced in the gap of the dimer antenna. Second, two line-like patterns of enhanced power loss are present at each side of the dimer antenna. These regions are located where the absolute value of the out-of-plane electric field is maximum. Consequently, the power loss within the GST is related to the enhanced electric field upon local switching. The self-consistent multiphysics simulation subsequently performed 500 ns after irradiation exhibits two regions left and right to the dimer antenna where crystallization occurs, and one in the gap, matching the regions of high power loss density (c.f. **Figure 4d**). The fact that the crystallization in the gap only has a small lateral extent, despite the power loss density being considerably higher in the gap, can be explained by the large thermal conductivity of the metallic antennas, yielding to a decreased temperature and hence less crystallization.

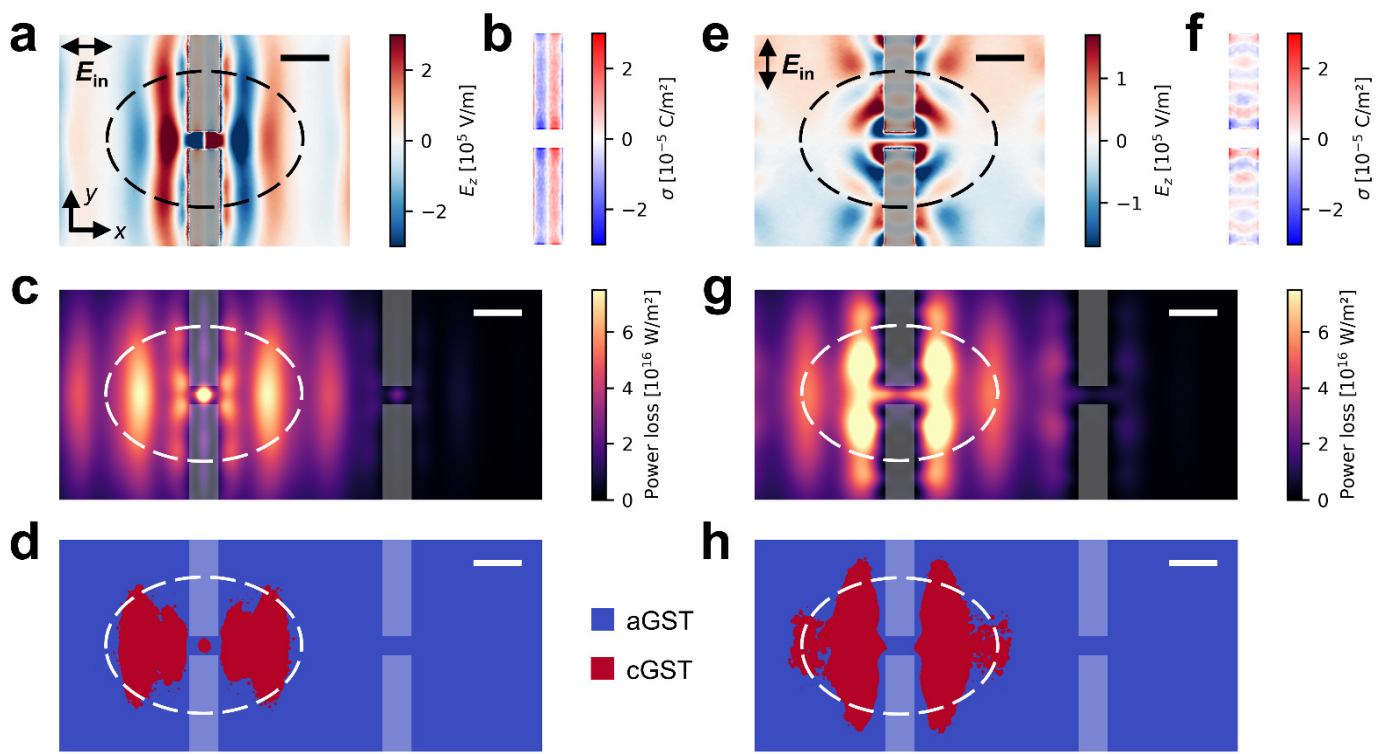

**Figure 4: Influence of polarization for the observed crystallization pattern.** The electric field of the laser excitation is polarized in two different directions: in *x*-direction (**a** - **d**) and in *y*-direction (**e** - **h**), as indicated in the upper-left of **a** and **e**. The approximate width of the experimental crystalline spot without antennas indicated by a dashed ellipse. The scale bars are 0.5 µm. **a), e)** The real part of *z* (out-of-plane) component of the electric field $E_z$ at the bottom of the antenna for different switching laser polarizations displays a wavy pattern induced by the dimer antenna. **b), f)** Surface charge density at the bottom of the dimer antenna **c), g)** Lateral cross-sections of the simulated power loss density through the top surface of the GST layer are displayed from above. **d), h)** Cross-sections through the top of the GST layer of the crystallization for the different polarization directions, acquired via multiphysics simulations.

If the polarization of the incident laser is rotated in the simulation, now aligning with the length of the antenna, the observable electric field distribution changes significantly as visualized in **Figure 4e**. The sign of $E_z$ alternates across the length of the dimer antenna, as is the case for the surface charges within the metallic dimers (see **Figure 4f**). The appearance of several minima and maxima can be explained by the oscillating charges in y-direction within the nanoantenna. The resulting calculated power loss within the GST layer shown in **Figure 4g** differs from the previously observed pattern in Figure 4c. Now, the power loss is enhanced directly next to the gap of the dimer antenna and extends towards the antenna ends.

The resulting crystallized area (c.f. **Figure 4h**) is in this case closer to the antenna and extends more in y-direction. Moreover, an additional region of crystallization appears on each side of the antenna, corresponding to the power loss shown in Figure 4g.

In summary, the electric field distribution induced upon locally addressing metallic nanoantennas as well as the thermal transport by the antennas strongly influences the observed crystallization behavior. The polarization of the applied switching laser tailors the surface charge density within those antennas and yields different crystallization behaviors dependent on the orientation of the antennas.

Finally, we investigate the time evolution of crystallization during laser irradiation of 500 ns. We focus on x-polarization perpendicular to the dimer antenna length. **Figure 5a** displays four distinct time steps of the crystallization obtained with multiphysics simulations for centrally addressed dimer antennas, i.e., from 100 ns to 500 ns, revealing the crystallization kinetics upon locally heating up amorphous GST with laser irradiation. The top panels show a lateral cross-section through the top of the GST layer, the bottom panels show a vertical cross-section through the center of the dimer antenna. No crystallization has occurred until 100 ns (i). At 200 ns (ii), some small crystalline nuclei have formed at the left and right side of the dimer antenna. Later at 300 ns (iii), these nuclei have grown and additional nuclei are visible. After 500 ns (iv), when the laser is turned off, the GST is crystallized in two regions next to the dimer antennas, resembling a butterfly-pattern when viewed from above. In accordance with the experiments and simulations discussed above, the lateral extent of the crystallization is largest close to the surface and shrinks with increasing depth (see bottom panel in Figure 5a (iv)). The finite crystallization depth in the simulations is around 40 nm. The time-dependent power loss at four exemplary positions (see colored dots in Figure 5a) is shown in **Figure 5b**. Up to 100 ns, the power loss remains constant, and the different values at each position result from the spatially inhomogeneous power loss distribution, exemplarily shown at 0 ns in Figure 4c. The strong increases of the power losses at around 150 ns for position 3 and at around 300 ns at positions 2 and 4 are associated with crystallization of the amorphous GST, resulting in stronger light absorption due to the increased losses within crystalline GST (see permittivity of GST in the visible in **Supplementary Note 4**). This phenomenon is also visible in the spatial power loss distribution at 300 ns in **Figure 5c**.

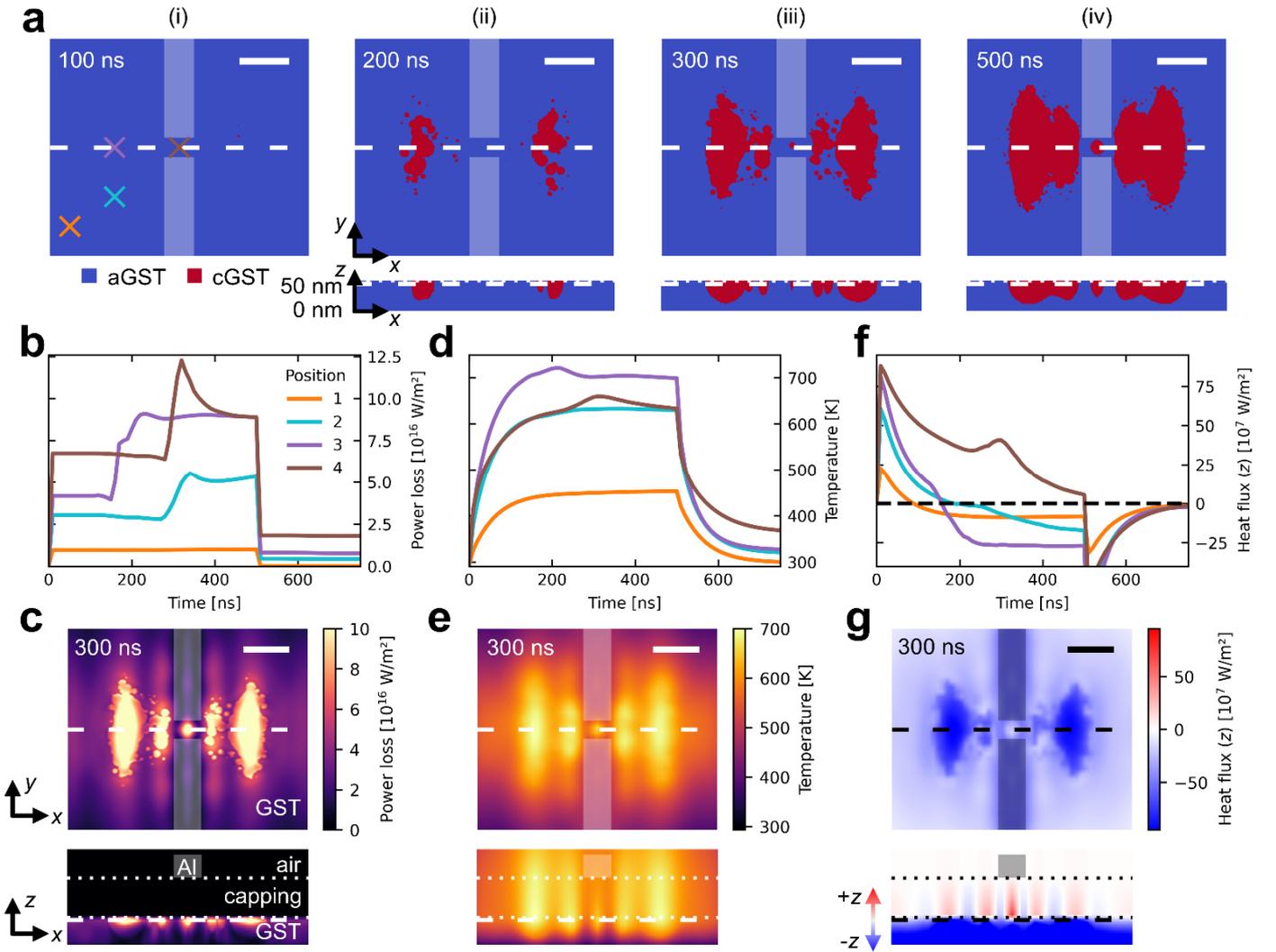

**Figure 5: Time evolution of the simulated thermodynamic quantities upon crystallization.** The excitation corresponds to center addressing with *x*-polarized light and is turned off after 500 ns. **a)** Simulated crystallization of the GST layer after 100, 200, 300 and 500 ns. Top panels: *x-y*-plane on top of the PCM (view from above). Bottom panels: cross-section (*x-z*-plane) perpendicular to the dimer antenna and through its center (view from the side). Note that the image is stretched in *z*-direction due to the low thickness of the PCM. **b), d), f)** Simulated power loss density (**b**), temperature (**d**) and heat flux in *z*-direction (**f**) at the four positions indicated in the leftmost panel of **a** as a function of simulation time. **c), e), g)** Spatial cross-sections of the power loss density (**c**), temperature (**e**) and heat flux (**g**) at 300 ns, the bright regions indicate crystallized GST featuring stronger power loss and absorption. The location of the antennas above is indicated by semi-transparent rectangles. The dashed lines indicate the location of the corresponding cross-section. The scale bars are 0.5 µm.

The temperature at the respective positions (c.f. **Figure 5d**) rises with time while the incident laser is applied. For example, at position 3 (purple), the temperature rises above 700 K after 200 ns, and then mostly remains constant until the laser is turned off. The spatial temperature distribution after 300 ns is shown in **Figure 5e** and features a similar general pattern as the power loss density. In the vertical cross-section across the layer stack (bottom panel), the heat is mainly confined in the capping layer and the upper part of the GST layer. The large thermal conductivity of the substrate compared to the thermal conductivity of air confines the heat at the top. Tailoring the different layers and applying a second thermal barrier between the substrate and the GST layer might lead to more homogeneous crystallization across the layers.[50]

The simulated heat flux (see **Figure 5f**) at the top of the GST layer is positive at short time scales for the positions 1, 2, and 3, referring to heat transportation in positive z-direction toward the capping. For longer time scales, the heat flux turns negative at these positions, meaning that most of the heat is flowing towards the silicon substrate featuring a larger heat conductivity compared to air and the capping layer. The high heat flux in the substrate can also be observed in the spatial distribution of the heat flux (c.f. **Figure 5g,** bottom panel). In addition, the individual crystallized areas can be clearly distinguished (see top panel), due to an approximate doubling of thermal conductivity and resulting heat flux upon crystallization. Due to the increased power loss at positions where crystallization occurs, an increase in temperature would be expected. However, the increased power loss is balanced by the increased heat flux, resulting in only small and short fluctuations in temperature. For position 4, in the gap of the dimer antenna, the heat flux is positive for the whole 500 ns. The heat is transported toward the metallic antennas that feature a high thermal conductivity. Since the power loss is low in the PCM layer next to the gap, directly beneath the antennas, the temperature only increases slightly there and heat from position 4 can be dissipated there. As a result, the temperature at position 4 increases slower than e.g. at position 3, although the power loss at position 4 is higher most of the time.

**Discussion:**

In summary, we have investigated inhomogeneous crystallization of GST around infrared plasmonic nanoantennas upon visible laser irradiation. We demonstrated that measured infrared reflectance spectra of locally addressed dimer antennas require multiphysics simulations to reproduce the observable resonance shifts. With multiphysics simulations, we studied how the interaction of the incident laser with metallic nanoantennas leads to spatially inhomogeneous power absorption within the PCM layer. We found that the polarization of the incident switching laser plays a major role in the crystallization process. Moreover, the thermal heat conductivities of the adjacent layers and the antennas themselves strongly influence the crystallization. Time-resolved simulations of the power loss, temperature and crystallization shed light on complex phase transition kinetics usually not accessible within experimental setups.

We anticipate that the crystallization process can be modified further using other antenna geometries such as split-ring resonators or bow-tie antennas. Tailored crystallization spots of arbitrary shapes could be achieved by considering the interplay between metallic nanoantennas, customized layer stacks with suitable thermal properties, and arbitrary laser intensity profiles instead of the commonly utilized Gaussian profiles. Since crystallization can depend on the polarization of the switching laser, it might be possible to achieve metasurfaces showing polarization-dependent switching behavior by carefully tuning the geometry of the metallic nanostructures. We expect that the reamorphization process around nanoantennas will be influenced in a similar manner. Multiphysics simulations provide a unique possibility to explore the effects of these parameters.

Moreover, the plasmonic PCM $In_3SbTe_2$ gained much interest in the past years due to its metallic crystalline phase in the entire infrared spectral range[51],[52], enabling optical programming of plasmonic nanoantennas[53],[54], functional metasurfaces[55]–[57] and polariton optics[58],[59]. However, the crystallization behavior of IST is controversial, indicating that more detailed studies must be performed. A multiphysics model of IST has not been established yet, despite the rise of IST in various nanophotonic application areas. Finally, by incorporating more complex phenomena into the multiphysics simulations, such as crystal grains, anisotropy, temperature-dependent material properties and amorphization, and by extending the model to other PCMs such as the low-loss PCMs $Sb_2Se_3$ and $Sb_2S_3$[60],[61], an extensive multiphysics model could be constructed that accurately predicts the crystallization of PCMs in metasurfaces[1], photonic integrated circuits[62], and other electro-optical devices, significantly improving the design process of dynamic nanophotonic elements.

## Methods:

### Sample fabrication:

Direct current magnetron sputtering was employed to deposit the 50 nm amorphous $Ge_3Sb_2Te_6$ with a power of 20 W and a deposition rate of 0.15 nm/s on top of a 1x1 cm² silicon substrate with a thickness of 1 mm. Subsequently, radio frequency sputtering was utilized to deposit a 70 nm $ZnS:SiO_2$ capping layer with a power of 59 W and a deposition rate of 0.09 nm/s to prevent the PCM layer from oxidation. Afterward, 100 nm AR-P 639.04 PMMA resist was spin-coated with 3000 rpm on top, followed by 270 nm AR-P 679.04 PMMA resist with 4000 rpm. After each PMMA deposition step, the entire sample was heated for 5 minutes to 130 °C. The resist was patterned with an electron beam lithography setup eLine from Raith GmbH. An accelerating voltage of 10 kV, an aperture size of 10 μm, a beam current of 25.4 pA and a writing field of 80x80 μm² were utilized. After patterning the resist, the sample was developed for 52 s in a 3:1 mixture of deionized water and isopropyl alcohol and subsequently dried with nitrogen. Afterward, the sample was submerged for 20 s in pure isopropyl alcohol. For thermal evaporation, a Leica EM MED020 was employed with pure aluminum positioned in a tungsten boat and evaporated using 110 A. For the lift-off process, the sample was positioned in acetone for thirty minutes to dissolve the remaining resist material at 35 °C.

### Optical switching:

Optical crystallization of GST was performed with a home-built laser setup consisting of a laser diode with 660 nm wavelength, focused by a 10x objective with a numerical aperture of 0.25 onto the sample. The beam waist of the laser at the sample was 2.2 x 1.5 μm². The sample was moved with a Thorlabs closed-loop piezo controller (BCP303) for precise movements and Thorlabs NanoMax-TX stepper motors for coarse movements. The number of laser pulses at one spot were controlled by an external pulse generator (Keysight 3320A). The GST was crystallized with 21 laser pulses with a pulse length of 500 ns and a repetition rate of 200 Hz. Powers of 15.6 mW and 23.2 mW were employed for low-power and high-power crystallization, respectively.

### Sample characterization:

The scattering-type scanning near-field optical microscopy (s-SNOM) measurements were conducted with a neaspec neaSNOM equipped with a PtIr5-coated silicon tip (Arrow NCPt, NanoWorld) and operated in pseudo-heterodyne detection mode[63]. The sharp tip with a curvature radius of approximately 20 nm oscillated at about 260 kHz with an oscillation amplitude close to 80 nm, enabling simultaneous measurement of the topography using an atomic force microscopy setup and the optical near field signal consisting of amplitude and phase. The third-demodulation order optical amplitude images of the center-addressing case were acquired with a tunable quantum cascade laser (MIRcat-QT, Daylight Solutions) at 1000 $cm^{-1}$, the images for the center-addressing case were acquired at 2500 $cm^{-1}$ with an SI Stuttgart Instruments Alpha series post-amplified fiber-feedback optical parametric oscillator[64],[65], pumped with a picosecond mode-locked oscillator laser (M-PICO-LAB Yb, Montfort Laser) operated in pulsed mode. These frequencies are off-resonant compared to the antenna resonance (around 1700 $cm^{-1}$) and provide good contrast between aGST and cGST.

The Fourier-transform infrared reflectance (FTIR) spectra measurements were performed with a Bruker Optik HYPERION 2000 microscope connected to a Bruker Optik VERTEX 70 interferometer. The microscope unit was equipped with a 36x Schwarzschild objective, which has an angular spread ranging from about 8° to 30° and an average angle of approximately 20°. Linear polarizers in the beam path were used to select only the p-polarized component of the infrared light and knife-edge apertures were utilized to restrict the measurement area and minimize contributions of the surrounding. The spectra were averaged over 2000 scans with a resolution of 8 $cm^{-1}$ and referenced to a gold plate.

### Simulations:

The multiphysics simulations were performed with a custom, self-consistent multiphysics model, introduced by Meyer et al.[29] The laser excitation is represented by the electromagnetic field of a Gaussian beam that matches the experimental conditions (see Supplementary Note 2). The simulation process consists of three steps that are executed in a loop: First, the electromagnetic solver of CST Studio Suite (CST) is employed to acquire the power loss density. Second, this power loss is imported as a source for a thermal solver to calculate the temperature distribution after a specified time step. Third, the temperature is imported into a custom crystallization model to simulate the crystallization process during the specified time step. Both the electromagnetic and thermal solver are updated with the resulting crystallized volume to achieve self-consistency. The crystallization simulation is implemented in OpenCL via Python and is based on a phenomenological phase field model. See **Supplementary Note 5** for details.

The antennas and layer stack were defined according to the experimental parameters. All materials were assigned their respective thermal and electromagnetic properties, c.f. **Supplementary Note 6.** However, the crystallization simulation was performed with $Ge_2Sb_2Te_5$ instead of $Ge_3Sb_2Te_6$ due to lack of data. In the thermal simulations, thermal interfacial resistance is modeled by an additional 5 nm layer with reduced thermal conductivity above and below the PCM layer. The crystallization was simulated in a 5 µm x 2.2 µm domain, with a cubic voxel size of 2 nm, while the total domain sizes of the electromagnetic and thermal simulation were 10.5 µm x 7.8 µm and 7.5 µm x 4 µm, respectively. To approximate the experimental conditions, where the antennas are placed in an array, a total of 4 antennas were considered, of which the center two were inside the domain of the crystallization simulation. A time step of 10 ns and a total simulation time of 500 ns were chosen. Since the simulation of the 21 laser pulses used in the experiments would take several days with adequate temporal resolution, only a single pulse was simulated. On a sample without antennas, the simulated crystallization with these parameters had a similar extent as the crystallization observed in experiments, see Supplementary Note 2.

The reflectance spectra were calculated with the electromagnetic frequency domain solver of CST, using a tetrahedral mesh and an accuracy of $10^{-6}$. In lateral directions, periodic boundary conditions were applied. In vertical directions open boundary conditions were applied, with sufficient thicknesses of the air layer on top and the substrate on the bottom. For excitation, Floquet ports were chosen. The p-polarized part of the spectra was extracted, corresponding to polarization along the long axis of the dimer antenna, and they were averaged across integer incidence angles ranging from 8° to 30° according to the intensity distribution of the FTIR objective. The antennas were defined according to the experimental geometry shown in Figure 1, with the exception of the antenna length which was set to 1.1 µm, according to measurements with a light microscope. The lateral periods were 2 µm and 4.1 µm, respectively. For the naive case, the crystallized volume was approximated by an elliptical cylinder with a diameter corresponding to the experimentally observed spot size without antennas (see Supplementary Note 2) and the height being equal to the PCM thickness. For the spectra using the crystallization predicted by multiphysics simulations, the crystallized volume was approximated by one or several polygon curves that feature a vertical extent corresponding to the average crystallization depth.

## Acknowledgements:

The authors thank Maike Kreutz for the sputter deposition of the initial layer stack. L. S., L. C., L. J., T. T. and D. N. C. acknowledge support from the Deutsche Forschungsgemeinschaft through project No. 518913417. L. C. and T. T. acknowledge additional support from the Deutsche Forschungsgemeinschaft through SFB 917 "Nanoswitches". L. J. acknowledges additional support through the RWTH Graduate Support of RWTH Aachen University. D. N. C. acknowledges additional partial support from the Deutsche Forschungsgemeinschaft through a Heisenberg Fellowship (CH 407/13-1).

## Author contributions:

L. S. and L. C. contributed equally. L. S., L. C., T. T. and D. N. C. conceived the research idea and designed the research; Y. C. and L. C. performed the EBL, the thermal evaporation, the light microscopy, and the optical switching; L. J. captured the SNOM images; L. S. and S. M. performed the multiphysics simulations; L. S. and L. C. analyzed the data; M. W. provided the sputtering equipment and phase-change material expertise; all authors contributed to writing and editing the manuscript.

## Conflict of interest:

The authors declare no competing financial and/or commercial conflicts of interest.

## Data availability:

The data that support the findings of this study are available from the corresponding author upon reasonable request.

# Supporting Information for

**Understanding inhomogeneous crystallization dynamics of phase-change materials in the vicinity of metallic nanostructures**

***Authors:***

*Luis Schüler [a,b,\*,+], Lukas Conrads [b,+], Yingfan Chen [b], Lina Jäckering [b], Sebastian Meyer [b], Matthias Wuttig [b], Thomas Taubner [b], Dmitry N. Chigrin [a,b,#]*

[+] authors contributed equally

\* Email: schueler@physik.rwth-aachen.de
\# Email: chigrin@physik.rwth-aachen.de

**Affiliations:**
[a] AMO GmbH, Otto-Blumenthal-Straße 25, 52074 Aachen, Germany
[b] I. Institute of Physics (IA), RWTH Aachen University, Sommerfeldstraße 14, 52074 Aachen, Germany

**This PDF file includes:**

    **Supplementary Note 1: Electric field simulation of the dimer antenna**

    **Supplementary Note 2: Crystallization of GST without antennas**

    **Supplementary Note 3: Location of the cross-sections**

    **Supplementary Note 4: Dielectric function of GST in the visible range**

    **Supplementary Note 5: Details on the multiphysics model**

    **Supplementary Note 6: Material properties assumed in the simulations**

    **References**

## Supplementary Note 1: Electric field simulation of the dimer antenna

The real part of the electric field in *z*-direction (normal to the viewing plane) at the resonance wavelength of the dimer antenna (approximately 5.7 µm, c.f. Figure 2 in main text) is displayed in **Figure S1a**. The absolute value is highest at the ends of the individual rod antennas, with the sign alternating. Similarly, the electric field enhancement (absolute value with respect to the absolute value far away from the antenna) is highest at the ends of the individual rod antennas, and especially in the gap, see **Figure S1b**.

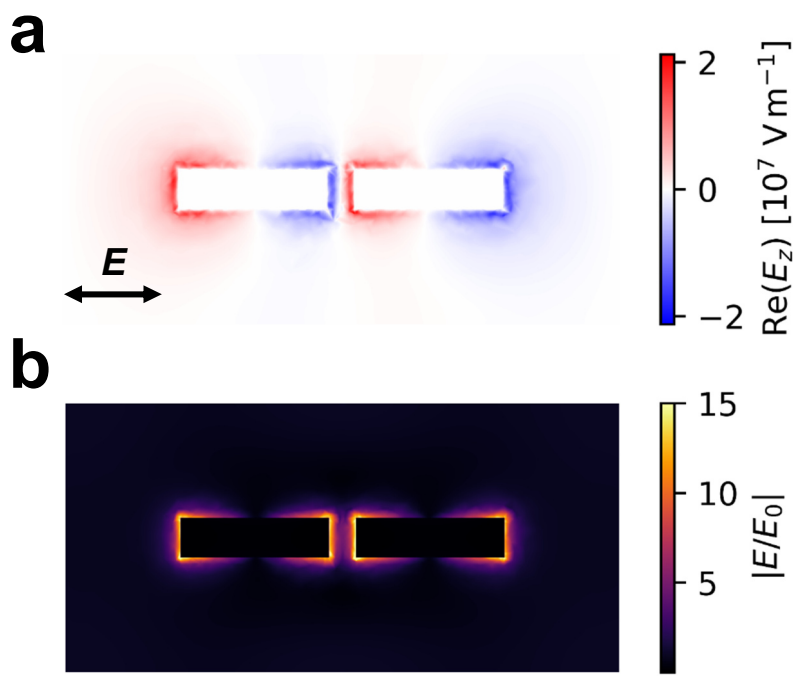

**Figure S1: Dipole resonance of the dimer antenna.** The polarization of the incoming electric field is along the long axis of the dimer antenna. **a)** Real part of the electric field in z-direction (normal to the viewing plane) at the resonance wavelength of the dimer antenna **b)**. Field enhancement (absolute value with respect to the absolute value far away from the antenna) at resonance.

## Supplementary Note 2: Crystallization of GST without antennas

In the experiment, 21 laser pulses were used. However, it would take unreasonably long to simulate these pulses. A constant time step of 10 ns, which provides adequate temporal resolution, would lead to a total simulation time of approximately 20 days. A dynamic adaptation of the time step, e.g. based on the rate of change of the temperature or the order parameter, could reduce the simulation time to several days, but would have to be carefully implemented so as not lose accuracy. Therefore, only one pulse was simulated. This is a reasonable approximation as a major part of the crystallized volume is already present after the first pulse, the subsequent pulses only cause the crystallized volume to expand slightly, especially for high laser powers. To ensure consistent results, a comparison of experimental and simulated laser powers was performed with a sample consisting of the same layer stack as described in the methods section, but without the dimer antennas. The GST was locally crystallized via laser irradiation and the spot sizes (major axis of the elliptical crystallized volume, as viewed from above) after irradiation with different powers and pulse lengths were measured. Afterward, multiphysics simulations were performed and the simulated spot sizes were extracted by fitting an ellipse. The experimentally measured and simulated spot sizes are shown in **Figure S2a**. Only one experimental spot was available for each laser power and pulse length. Each spot size was measured three times, and the standard deviation was used to estimate the measurement uncertainty. The simulated spot sizes match the measured values well for high laser powers. For low powers however, the simulations underestimate the spot sizes. This can be attributed mainly to the increased number of pulses in the experiment, but also to the limited availability of data for the crystallization of GST 326 (c.f. Supplementary Note 6) and to the ellipse fitting slightly underestimating the spot size. For small simulation times, no meaningful spot size could be determined. As an example, the time evolution of a simulated spot corresponding to a power of 15.6 mW is presented in **Figure S2b**. At 120 ns (first panel, top), no continuous crystalline area is present, and a spot size cannot be assigned. As the simulation time progresses, the crystallization forms a continuous spot, which is elliptic in lateral directions according to the intensity distribution of the simulated Gaussian beam and has a radial-dependent crystallization depth (bottom panel).

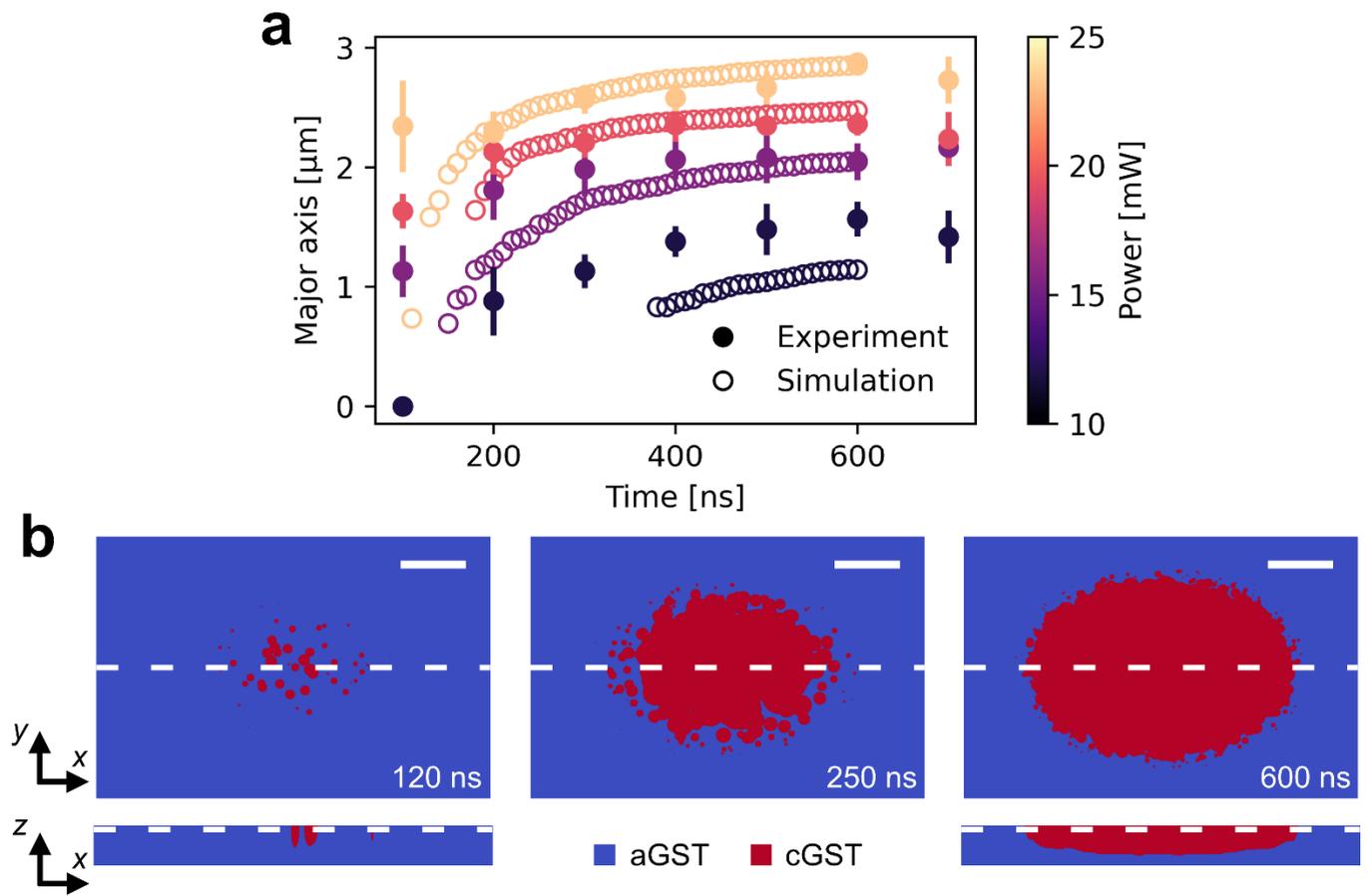

**Figure S2: Comparison of spot sizes. a)** Measured and simulated diameter (major axis) of the elliptical crystalline spots as a function of pulse length or simulation time after applying the calibration. **b)** Time evolution of a spot corresponding to an experimental power of 15.6 mW. Top: Cross section at the top of the PCM layer. Bottom: Cross-section along the major axis, showing the depth-dependence of the crystallization. The scale bars are 0.5 µm.

**Supplementary Note 3: Location of the cross-sections**

The cross-sections of the electrodynamic and thermodynamic quantities presented in the main text are extracted at two locations, indicated in **Figure S3**. The electric field and the surface charge density are extracted at the bottom of the dimer antenna, since their features are defined by the interaction of the antenna with the laser. The power loss density and heat flux are directly affected by crystallization, which only occurs inside the GST layer. Thus, these three quantities are also extracted there. The top of the GST layer is chosen as the location of the cross-sections since the lateral extent of the crystallization is highest there and the power loss density in the GST layer is also highest at the top, see Figure 5 in the main text.

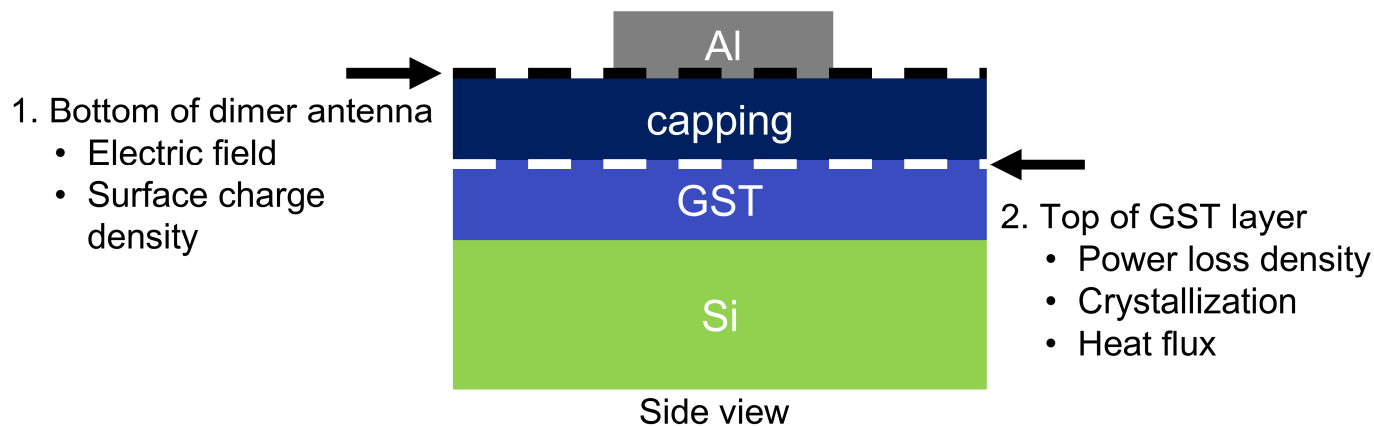

**Figure S3: Sketch of the cross-section locations.** There are two different locations for the cross sections, depending on which quantity is considered.

**Supplementary Note 4: Dielectric function of GST in the visible range**

The dielectric function of GST 326 in the visible range is shown in **Figure S4**. At the switching wavelength of 660 nm, the imaginary part of the permittivity of crystalline GST is more than twice as high as that of amorphous GST.

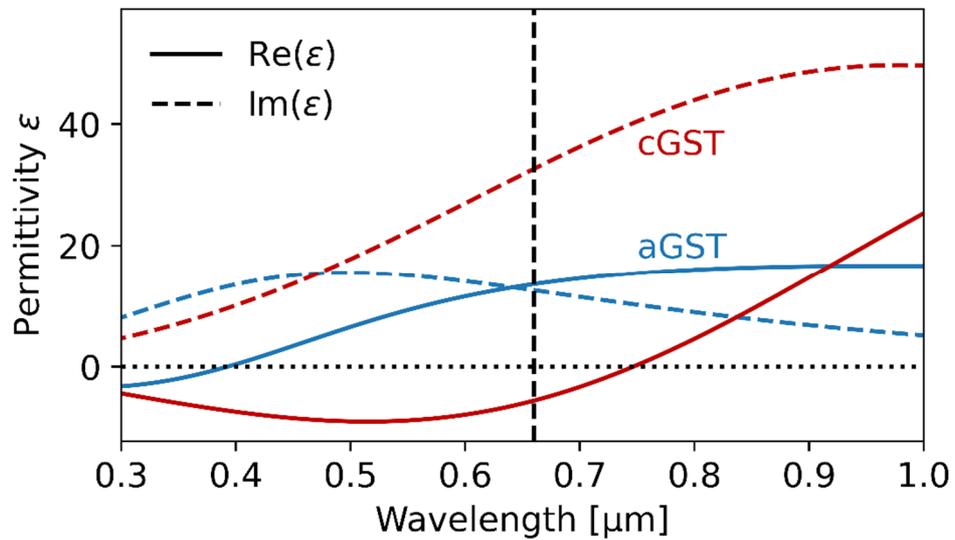

**Figure S4: Dielectric function of GST 326 in the visible range.** The wavelength of the switching laser (660 nm) is highlighted with a dashed line.

**Supplementary Note 5: Details on the multiphysics model**

The multiphysics simulations were performed with a custom, self-consistent multiphysics model, introduced by Meyer et al.[1] The laser excitation is represented by the electromagnetic field of a Gaussian beam that matches the experimental conditions (see Supplementary Note 1). The simulation process consists of three steps, c.f. **Figure S5**: First, the electromagnetic time-domain solver of CST Studio Suite (CST) is employed to acquire the power loss density using the finite integration technique for a hexahedral mesh with an accuracy of $10^{-6}$. Second, this power loss density is imported as a source for the thermal transient solver of CST to calculate the temperature distribution after a specified time step, again using the finite integration technique for a hexahedral mesh and an accuracy of $10^{-6}$. Third, the temperature is imported into a custom crystallization model to simulate the crystallization process during the specified time step. Both the electromagnetic and thermal solver are updated with the resulting crystallized volume to achieve self-consistency. These 3 steps are repeated consecutively until the set simulation time is reached. The crystallization simulation is implemented in OpenCL via Python and is based on a phenomenological phase field model, where an order parameter can assume values between 0 and 1, where 0 represents the amorphous phase and 1 the crystalline phase of the PCM. The evolution of the order parameter is described by the Allen-Cahn equation.[2]

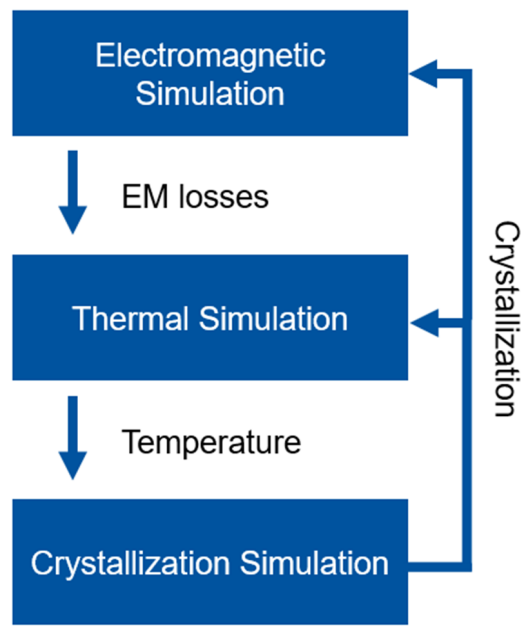

**Figure S5: Sketch of the multiphysics simulation loop.** The simulated electromagnetic losses serve as a source field to calculate the temperature, which in turn serves as the basis for the crystallization simulation. The resulting crystallization distribution is self-consistently coupled back to the electromagnetic and thermal solvers.

## Supplementary Note 6: Material properties assumed in the simulations

The assumed dielectric and thermal parameters of the different materials (permittivity at the wavelength of the switching laser, 660 nm) are given in **Table S1**.

**Table S1: Assumed dielectric and thermal material parameters.**

|  | Al | ZnS:SiO$_2$ | aGST | cGST | Si |
|---|---|---|---|---|---|
| Permittivity | -46.38 + $i$18 | 4.56 | 13.57 + $i$12.61 | -5.63 + $i$32.54 | 14.7 |
| Thermal conductivity [W K$^{-1}$ m$^{-1}$] | 237 | 0.55 | 0.195 | 0.5 | 148 |
| Specific heat [J K$^{-1}$ kg$^{-1}$] | 900 | 530 | 260 | 260 | 700 |

Due to lack of data on the crystallization dynamics of Ge$_3$Sb$_2$Te$_6$ (GST 326), the crystallization was simulated with the parameters of Ge$_2$Sb$_2$Te$_5$ (GST 225). Both materials have very similar properties, and the model for GST 225 has been successfully employed before to simulate the crystallization of GST 326[3],[4]. As detailed by Meyer et al.,[1] the nucleation rate is calculated with classical nucleation theory, and the growth velocity is given by the Arrhenius model[5] for temperatures below the glass transition temperature and by the Wilson Frenkel mode[6],[7] for temperatures above the glass transition temperature, ensuring a smooth transition at the glass transition temperature. The assumed temperature dependences of the nucleation rate and the crystal growth velocity are shown in **Figure S6**.

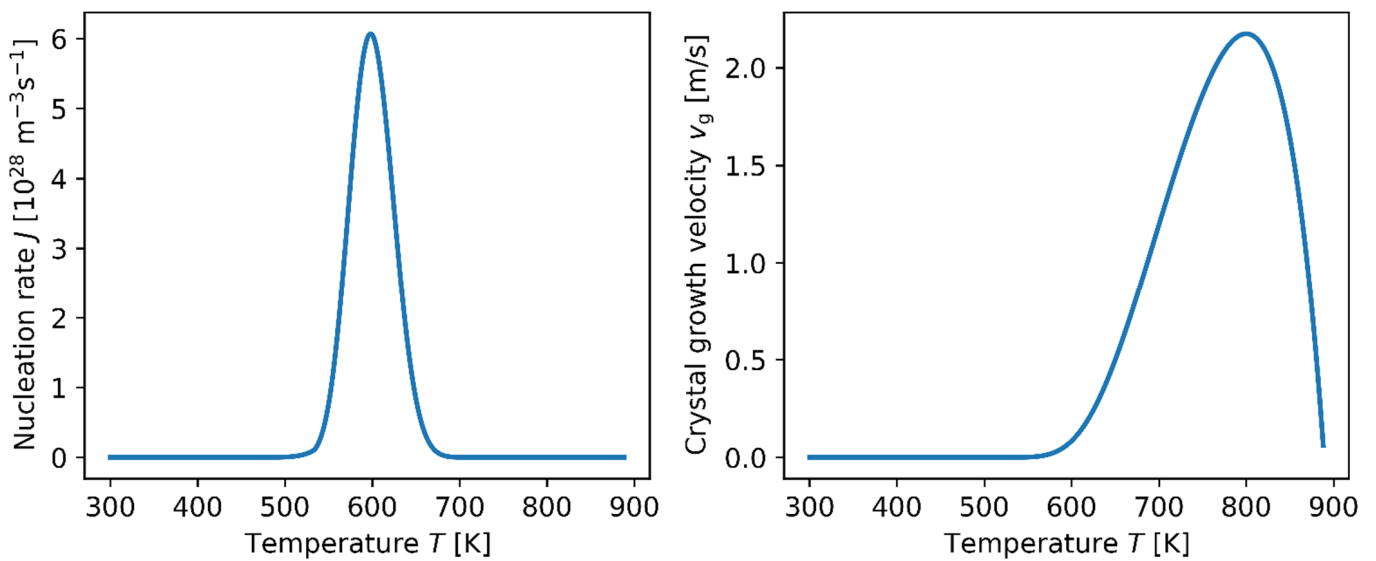

**Figure S6: Temperature-dependent nucleation rate and crystal growth velocity of GST 225.**

In particular, the nucleation rate is given by

$$J = 4\frac{j_\text{f}}{v_\text{m}} n_\text{c}^{2/3} \sqrt{\frac{\Delta G}{3\pi n_\text{c}^2\, k_\text{B} T}} \exp\left(-\frac{\Delta G_\text{c}}{k_\text{B} T}\right),$$

Where $j_\text{f}$ is the molecular jumping frequency, $v_\text{m}$ is the volume per monomer, $n_\text{c}$ is the number of molecules in a critical nucleus, $\Delta G$ is the free energy difference, $k_\text{B}$ is the Boltzmann constant, and $\Delta G_\text{c}$ is the critical nucleation free energy.[1],[5] The molecular jumping frequency is given by the Stokes-Einstein relation

$$j_f = \frac{k_\text{B} T}{3\pi \lambda^3 \eta},$$

where $\lambda$ is the molecular jump distance and $\eta$ is the temperature-dependent viscosity.[5] The number of molecules in a critical nucleus $n_\text{c}$ is defined by

$$n_\text{c} = \frac{4}{3}\pi \frac{r_\text{c}^3}{v_\text{m}},$$

where $r_\text{c} = 2\gamma v_\text{m}/\Delta G$ is the critical radius of the nucleus, with the interfacial energy $\gamma$.[5] The free energy difference $\Delta G$ is given by the Thompson-Spaepen equation

$$\Delta G = v_\text{m} L \frac{T_\text{m} - T}{T_\text{m}} \frac{2T}{T_\text{m} + T},$$

where $L$ is the latent heat per unit volume and $T_\text{m}$ is the melting temperature.[8] The critical nucleation free energy $\Delta G_\text{c}$ is given by[5]

$$\Delta G_\text{c} = \frac{16}{3}\pi \frac{\gamma^3 v_\text{m}^2}{\Delta G^2}.$$

The viscosity $\eta$ is given by a combination of the Arrhenius model[5] for temperatures below the glass transition temperature $T_\text{glass}$ and the Mauro-Yue-Ellison-Gupta-Allan (MYEGA) model[9] for temperatures above $T_\text{glass}$:

$$\eta = \eta_\infty \exp\left(\frac{E_\text{a}}{k_\text{B} T}\right), \quad (T \leq T_\text{glass}),$$

$$\log_{10} \eta = \log_{10} \eta_\infty + (12 - \log_{10} \eta_\infty)\frac{T_\text{g}}{T} \exp\left[\left(\frac{m}{12 - \log_{10} \eta_\infty} - 1\right)\left(\frac{T_\text{g}}{T} - 1\right)\right], \quad (T > T_\text{glass}),$$

where $\eta_\infty$ is the viscosity limit at high temperatures, $E_\text{a}$ is the activation energy, $T_\text{g}$ is the temperature where $\eta = 10^{12}$ Pa, and $m$ is the fragility that characterizes the slope of the viscosity at $T_\text{g}$. At $T_\text{glass}$, a smooth transition between both models is ensured. Similarly, as mentioned above, the crystal growth velocity is given by the Arrhenius model[5] for temperatures below the glass transition temperature and by the Wilson Frenkel model[2],[7] for temperatures above the glass transition temperature, again ensuring a smooth transition at the glass transition temperature $T_\text{glass}$:

$$v_\text{g} = v_\infty \exp\left(\frac{E_\text{a}}{k_\text{B} T}\right), \quad (T \leq T_\text{glass}),$$

$$v_\text{g} = \frac{4 r_\text{a} k_\text{B} T}{3\pi \lambda^2 r_\text{h} \eta}\left(1 - \exp\left(-\frac{\Delta G}{k_\text{B} T}\right)\right), \quad (T > T_\text{glass}),$$

where $v_\infty$ is the velocity limit at high temperatures, $r_\text{a}$ is the atomic radius of a monomer, and $r_\text{h}$ is the hydrodynamic radius of a monomer.[1] The values of the parameters are given in **Table S2**.

**Table S2: Assumed parameters for the crystallization model.**[1],[5],[10],[11]

| Parameter | Value |
|---|---|
| Molecular jump distance $\lambda$[5] | 2.99 Å |
| Interfacial energy $\gamma$ | 0.06 J/m$^2$ |
| Volume per monomer $v_m$[5] | 2.9 × 10$^{-28}$ m$^3$ |
| Latent heat per unit volume $L$[5] | 6.25 × 10$^8$ J/m$^3$ |
| Melting temperature $T_m$ | 889 K |
| Glass transition temperature $T_{glass}$[10] | 534 K |
| Viscosity limit at high temperatures $\eta_\infty$[10] | 0.012 Pa s |
| Activation energy $E_a$ | 2.3 eV |
| Temperature $T_g$ where $\eta = 10^{12}$ Pa[10] | 472 K |
| Fragility $m$[10] | 140 |
| Velocity limit at high temperatures $v_\infty$ | 7.1 × 10$^{17}$ m/s |
| Atomic radius of a monomer $r_a$[10] | 0.1365 nm |
| Hydrodynamic radius of a monomer $r_h$[10] | 0.1365 nm |